\documentclass[aps,prb,twocolumn,superscriptaddress,showpacs]{revtex4}
\usepackage{graphicx}
\begin{document}
\hyphenation{TiOCl TiOBr}
\title{One-dimensional versus two-dimensional correlation effects in the oxyhalides TiOCl and TiOBr}

\author{M.~Hoinkis}
\affiliation{Experimentelle Physik 4 and R\"ontgen Research Center
for Complex Material Systems, Universit\"at W\"urzburg, D-97074
W\"urzburg, Germany} \affiliation{Experimentalphysik II,
Universit\"at Augsburg, D-86135 Augsburg, Germany}

\author{M.~Sing}
\author{S.~Glawion}
\affiliation{Experimentelle Physik 4, Universit\"at W\"urzburg,
D-97074 W\"urzburg, Germany}

\author{L.~Pisani}
\author{R.~Valent\'i}
\affiliation{Institut f\"ur Theoretische Physik, Universit\"at
Frankfurt, D-60054 Frankfurt, Germany}

\author{S. van Smaalen}
\affiliation{Laboratory of Crystallography, University of
Bayreuth, D-95440 Bayreuth, Germany}

\author{M.~Klemm}
\author{S.~Horn}
\affiliation{Experimentalphysik II, Universit\"at Augsburg,
D-86135 Augsburg, Germany}

\author{R.~Claessen}
\email[]{claessen@physik.uni-wuerzburg.de}
\affiliation{Experimentelle Physik 4 and R\"ontgen Research Center
for Complex Material Systems, Universit\"at W\"urzburg, D-97074
W\"urzburg, Germany}

\date{\today}

\begin{abstract}
We have performed a comparative study of the electronic structures
of the spin-Peierls systems TiOCl and TiOBr by means of
photoemission spectroscopy and density functional calculations.
While the overall electronic structure of these isostructural
compounds is qualitatively similar, the bromide appears to be less
one-dimensional. We present a quantitative analysis of the
experimental dispersions in terms of exchange constant $J$ and
hopping integral $t$ as well as a discussion of the qualitative
spectral features. From that we conclude that despite the
one-dimensional physics triggering the ground state in both
compounds a proper description of the normal state electronic
structure has to take into account the anisotropic frustrated
interchain interactions on the underlying triangular lattice.
\end{abstract}

\maketitle

\section{Introduction}

Based on magnetic measurements and LDA+U calculations, the layered
Mott insulator TiOCl was recently proposed to be electronically
and magnetically quasi-one-dimensional and an unusual spin-Peierls
compound.\cite{Seidel03} Indeed, a true spin-Peierls ground state
was identified for TiOCl as well as for the isostructural TiOBr by
revealing the lattice dimerization in x-ray diffraction
measurements.\cite{Shaz05,Palatinus05} For a conventional
spin-Peierls system one would expect only one phase transition,
however, TiOCl and TiOBr possess two successive phase transitions
at temperatures $T_{c_1}=67$\,K, $T_{c_2}=91$\,K and
$T_{c_1}=28$\,K, $T_{c_2}=47$\,K, respectively. In the
intermediate phase an incommensurate superstructure was found that
develops below $T_{c_2}$.\cite{Smaalen05,Krimmel06} In a first
order transition at $T_{c_1}$ this order locks in, resulting in
the commensurate order of the spin-Peierls ground
state.\cite{Ruckamp05,Smaalen05,Krimmel06} Also in contradiction
to a canonical spin-Peierls scenario, both compounds display
marked fluctuation effects in an extended temperature regime above
$T_{c_2}$. For TiOCl, this was seen in magnetic
resonance,\cite{Imai03,Kataev03} Raman\cite{Lemmens04} and
infrared spectroscopy,\cite{Caimi04} and specific heat
measurements.\cite{Hemberger05} For TiOBr, the importance of
fluctuations was inferred from the infrared optical
properties.\cite{Caimi04b} The origin of these fluctuations is
still unclear but fluctuating orbital degrees of freedom can be
ruled out due to the high energy of the lowest crystal field
excitation in the $t_{2g}$
manifold.\cite{Ruckamp05,Hoinkis05,Zakharov06} From photoemission
spectroscopy on TiOCl in connection with electronic structure
calculations we argued in a previous study that electronic
correlations and/or spin-Peierls fluctuations might play an
important role in this system.\cite{Hoinkis05} The detailed
behavior of the quasi-one-dimensional dispersions, however, is
still not understood.

In this study we compare the electronic structures of TiOBr and
TiOCl well above the transition temperatures $T_{c_1}$ and
$T_{c_2}$. We present angle-integrated and angle-resolved
photoemission (ARPES) data as well as complementary density
functional calculations. While TiOBr is less one-dimensional than
TiOCl we conclude from a detailed analysis of the experimental
dispersions as well as a discussion of the qualitative spectral
features that probably a Hubbard-like model including the
anisotropic frustrated interchain interactions on the underlying
triangular lattice is best suited to capture the relevant physics
dominating the single-particle excitation spectra of these
compounds.

\section{Technical Details}
Single crystals of TiOCl were prepared by chemical vapor transport
from TiCl$_3$ and TiO$_2$.\cite{Schaefer58} TiOBr crystals were
synthesized in a similar way from TiO$_2$ with a 40\% excess of Ti
and TiBr$_4$.\cite{Palatinus05,Schaefer58}

Photoemission spectroscopy (PES) was performed using He\,\textsc{i}
radiation (21.22\,eV) and an Omicron EA~125~HR electron energy
analyzer. Fresh surfaces were prepared by \emph{in situ} crystal
cleavage with adhesive tape. The resulting surfaces were clean and
atomically long-range ordered as evidenced by x-ray induced
photoemission (XPS) and low-energy electron diffraction (LEED),
respectively. Since both TiOCl and TiOBr are insulators sample
charging was reduced by measuring at elevated temperatures,
\emph{i.e.}, at 325\,K in the case of TiOCl and at 360\,K in the
case of TiOBr. For TiOCl, we found from systematic temperature
variations that the charging is almost negligible at and above
$\approx$\,370\,K. In this situation the maximum of the Ti~$3d$ peak
(see below) is located 1.45\,eV below the experimental chemical
potential $\mu_{\mathrm{exp}}$, which corresponds to the Fermi edge
position of a silver foil. The TiOCl spectra were aligned
accordingly. The energy position of the Ti~$3d$ peak in TiOBr
depends on the temperature in a very similar way. It was aligned to
the same value of 1.45\,eV below $\mu_{\mathrm{exp}}$. The energy
resolution amounted to 60\,meV for the TiOCl and 110\,meV for the
TiOBr measurements. The angular acceptance was $\pm1^\circ$ in both
cases.

The photoemission experiments are complemented by calculations of
the electronic structure of TiOBr and TiOCl within the \emph{ab
initio} density functional theory in the generalized gradient
approximation with inclusion of on-site Coulomb correlations
(GGA+U)\cite{Perdew96,Anisimov93} using the full-potential
linearized augmented plane-wave code WIEN2k.\cite{wien2k} The GGA+U
calculations, which require a spin-polarized starting configuration,
were performed by considering a ferromagnetic alignment of the Ti
spins. For both systems a value of $RK_{\mathrm{max}}=7$ was used
and  $40$k irreducible points were considered for Brillouin zone
integrations. The onsite Coulomb repulsion $U$ and onsite exchange
$J_0$ were chosen to be $3.3$\,eV and $1$\,eV,
respectively.\cite{Seidel03,Saha-Dasgupta04}

\section{Results and Discussion}

\begin{figure}
\includegraphics[width=8.2cm]{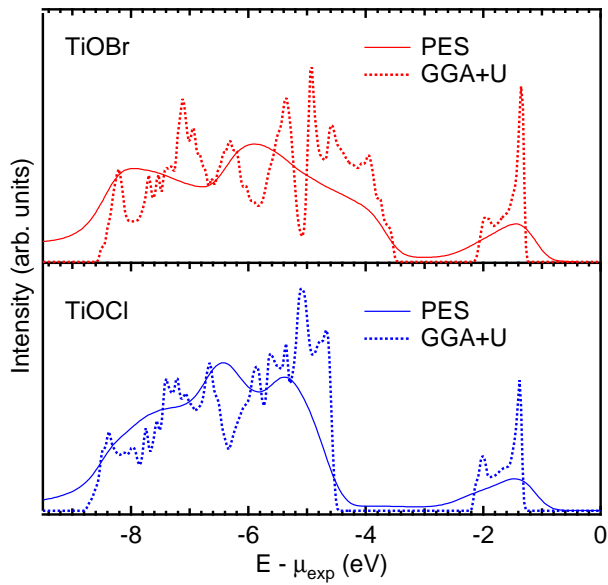}
\caption{\label{Figure1}(Color online) Angle-integrated
photoemission spectra and GGA+U densities of states of TiOBr and
TiOCl.}
\end{figure}

In Fig.~\ref{Figure1} angle-integrated photoemission spectra are
compared to the GGA+U densities of states. The calculations yield a
clear separation of the states with predominant Ti~$3d$ character
centered around $-1.5$\,eV from the ones which are mainly derived
from the O and Br/Cl~$p$ levels. The latter are found between
$-8.5$\,eV and $-3.5$\,eV for TiOBr and between $-9$\,eV and
$-4.5$\,eV for TiOCl. The photoemission spectra clearly show this
partition as well. Also the significantly smaller separation of the
Ti~$3d$ states from the rest of the valence band which is found in
the calculations for TiOBr is observed in photoemission.

However, neither the exact shape nor the width of the Ti~$3d$
spectral weight are reproduced by our calculations. As it was
pointed out in an earlier publication,\cite{Hoinkis05} this is most
likely due to pronounced electronic correlations and/or spin-Peierls
fluctuations which are beyond the scope of the GGA+U method. A
better match between the calculated and the experimental spectra is
found with respect to the width of the $p$ bands. Due to matrix
element effects a more detailed agreement of the spectra with the
bare densities of states cannot be expected.

\begin{figure}
\includegraphics[width=8.2cm]{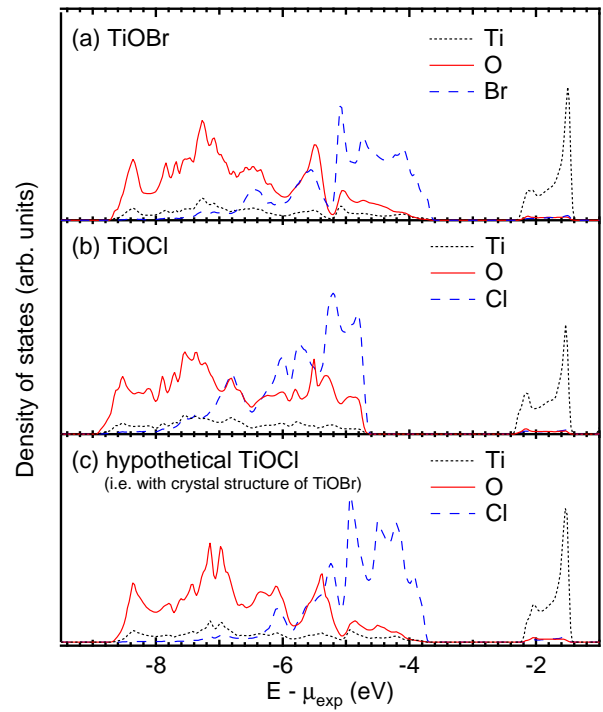}
\caption{\label{Figure2}(Color online) Atomically resolved GGA+U
densities of states of (a) TiOBr, (b) TiOCl, and (c) a hypothetical
compound with the crystal structure of TiOBr but with Br substituted
by Cl.}
\end{figure}

In order to address the question whether the smaller separation
between the manifold of O/Cl(Br) and the Ti $3d$~states in the Br
compared to the Cl system is caused by the change in chemistry or
rather related to the structural expansion we performed a third
calculation. There we assumed a hypothetical compound with the
crystal structure of TiOBr but with Br substituted by Cl. In
Fig.~\ref{Figure2} the atomically resolved densities of states for
all three systems is presented. The decomposition into Ti states
near the chemical potential on the one hand and O and Cl(Br) states
at higher binding energies on the other becomes very clear in this
plot. By comparison it is easily seen that the separation of the Ti
states is smaller in the Br system mainly due to the expanded unit
cell of TiOBr.

\begin{figure}
\includegraphics[width=7.2cm]{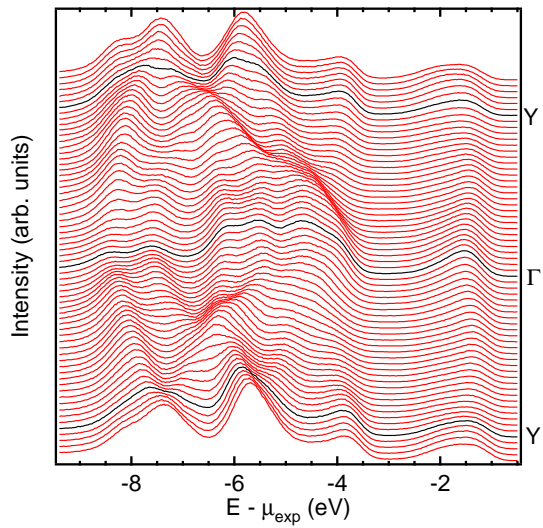}
\caption{\label{Figure3}(Color online) EDCs of TiOBr along the
crystallographic $b$~axis, corresponding to the Y$\Gamma$Y line in
the orthorhombic Brillouin zone.}
\end{figure}

Turning to our angle-resolved photoemission data, we display
energy distribution curves (EDCs) of TiOBr in a broad energy range
along the crystallographic $b$~direction in Fig.~\ref{Figure3}.
What immediately stands out are the well-pronounced dispersions,
especially in the Br~$4p$/O~$2p$ part of the spectra. The
dispersions are clearly symmetric with respect to the $\Gamma$
point reflecting atomic long range order and thus can be taken as
a further proof of the good surface quality.

\begin{figure}
\includegraphics[width=8.2cm]{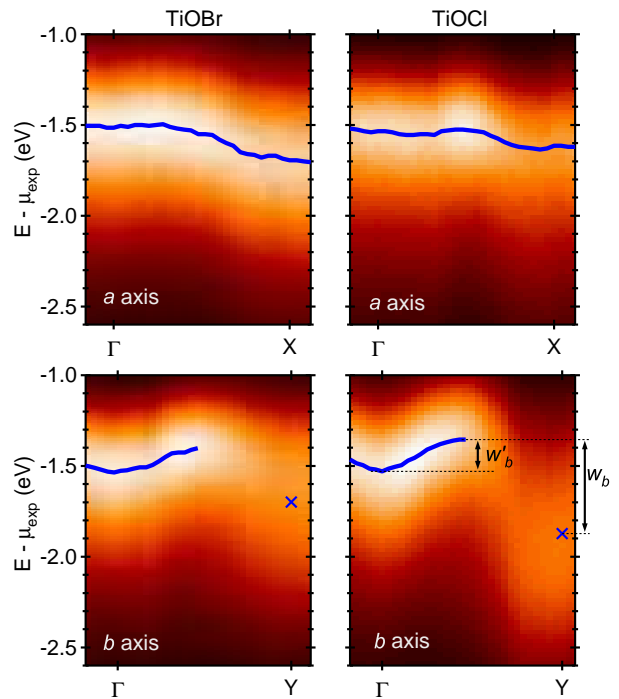}
\caption{\label{Figure4}(Color online) ARPES intensity plots
$I(\textbf{k},E)$ of TiOBr and TiOCl along the crystallographic
axes $a$ and $b$, corresponding to the X$\Gamma$X and Y$\Gamma$Y
lines in the orthorhombic Brillouin zone. Blue lines indicate the
peak positions of the EDCs. The blue crosses mark the first moment
of the EDCs at the Y points.}
\end{figure}

In Fig.~\ref{Figure4} the Ti~$3d$ spectral weight distributions of
TiOBr and TiOCl are compared. The four panels show ARPES intensity
plots $I(\textbf{k},E)$ along the crystallographic axes $a$ and $b$.
As seen, small but clear dispersions are discernible in all four
panels. These are quantitatively identified by lines and markers.
The lines indicate EDC peak maxima obtained by a fitting procedure.
For both compounds, this works well along the $a$ direction and in
the first half of the path from $\Gamma$ to Y, \emph{i.~e.} along
$b$ in direct space. In the second half, however, the EDCs lose
their single peak-like shape as an additional feature at higher
binding energies appears and thereby the spectra gradually change
their form to end up in a broad hump at the Y point.\cite{Hoinkis05}
Thus a fit with a single peak is not adequate here. Instead, we take
the first order moment of the EDCs at the Y point, indicated by
crosses in Fig.~\ref{Figure4}.

Following first the dispersions along the $a$ axes, \emph{i.~e.}
from $\Gamma$ to X in reciprocal space, both compounds display a
qualitatively similar behavior: Initially the energy changes only
slightly before it reaches a maximum, and then shifts downwards
until the Brillouin zone edge. Also along the crystallographic $b$
direction, the dispersions show a qualitative resemblance in the two
compounds. Starting from $\Gamma$, a single peak disperses upwards,
with the shape of the EDCs remaining essentially unaltered until
$\frac{1}{2}\Gamma$Y. As already pointed out, subsequently the
spectral changes get more complex. In any case the overall
dispersion bends downwards reaching a minimum at the Y point which
lies further below the one at $\Gamma$. We emphasize once more that
while at first glance the dispersions from the intensity plots might
appear to be similar along the $a$ and $b$ directions they are
qualitatively different if one looks at the spectral shapes of the
corresponding EDCs.

\begin{table}
\caption{\label{Table1}Dispersions $w_a$, $w_b$, and $w_{b'}$ of the
Ti~$3d$ weight in TiOBr and TiOCl, measured from the maximum to the
minimum energy of the peak for the direction~$a$, and from the
maximum energy of the peak to the first moment of the Y point EDC
for the direction~$b$, respectively.}
\begin{ruledtabular}
\begin{tabular}{lrr}
width & TiOBr & TiOCl\\
\hline
$w_a$&    0.27(3)\,eV & 0.12(3)\,eV\\
$w_b$&    0.26(5)\,eV & 0.47(5)\,eV\\
$w_b^{\prime}$& 0.13(1)\,eV & 0.17(1)\,eV\\

\end{tabular}
\end{ruledtabular}
\end{table}

For a quantitative comparison of the electronic dispersions between
the two compounds we list the parameters $w_a$, $w_b$, and
$w_b^{\prime}$ in Table~\ref{Table1}. The unprimed quantities refer
to the overall dispersion widths while $w_b^{\prime}$ measures the
width of only the inner part of the dispersion in the region from
$\Gamma$ to about $\frac{1}{2}\Gamma$Y. These widths are determined
either from the difference of the maximal and minimal peak energies
along the corresponding paths in $k$ space as obtained by the
fitting procedure (see lines in Fig.~\ref{Figure4}), or the
difference between the maximum peak energy and the first order
moment at the Y point in case of $w_b$ (see markers in
Fig.~\ref{Figure4}). The errors indicated reflect the scatter from
several samples and measurements.

As is immediately read off from Table~\ref{Table1} the overall
dispersion width along $b$ is significantly smaller in TiOBr with
respect to TiOCl while it is the other way round regarding the $a$
axis. Given that both systems and in particular their electronic
structures are governed by the same physics we hence conclude that
the anisotropy of TiOBr is less pronounced than in TiOCl. This is in
line with the trend in the relevant hopping integrals as derived
from downfolding LDA+U results as well as with the fact that the
Bonner-Fisher curve indicative for 1D Heisenberg chains does not
provide a good fit for the high-temperature magnetic susceptibility
of TiOBr in contrast to TiOCl.\cite{Lemmens05,footnote-ideal2D}

At this stage we resort to the central result of our previous study
on TiOCl\cite{Hoinkis05} where among various approaches to the $k$
resolved electronic structure we identified the spectral function of
the single-band 1D Hubbard model as the most promising starting
point in order to get further insight into the microscopic physics
behind our experimental data. We recall that the most obvious
shortcomings of the model calculations were the lack of any evidence
for spin-charge separation and the absence of the so-called shadow
band in the ARPES spectra although for the relevant parameter regime
it should be clearly observable.\cite{Benthien05} As possible
explanations for the discrepancies between theory and experiment we
invoked multiorbital effects and/or spin-Peierls fluctuations. A
comparison between the isostructural compounds TiOBr and TiOCl now
opens the possibility to further explore the reasons for this
discrepancy between 1D Hubbard model prediction and experiment. For
this purpose we focus on the central part of the dispersion along
the $b$ axis marked by lines in Fig.~\ref{Figure4}. Within the
single-band 1D Hubbard model this part corresponds to the ($\omega,
k$) region of the spinon and holon branches (see Fig.~7 in
Ref.~\onlinecite{Hoinkis05}). Hence, depending on the dominant
character of the experimental dispersion  its width $w_b^{\prime}$
should either scale with the exchange constant $J$ whose value can
be extracted from magnetic susceptibility
measurements\cite{Seidel03,Ruckamp05} or the hopping integral $t$.
The latter either can be inferred from the Hubbard model
perturbation expression for the exchange constant, $J=4t^2/U$, and
thus should scale as $t\propto \sqrt{J}$ or it can be deduced from
an appropriate downfolding procedure of LDA+U band
calculations.\cite{Lemmens05} An account of these quantities is
given in Table~\ref{Table2}. The values of $J$ for TiOBr and TiOCl
differ by $-45$\% (with respect to the value in TiOCl), whereas the
experimental dispersion width is smaller in TiOBr by only $23$\%.

\begin{table}
\caption{\label{Table2}Comparison of TiOBr and TiOCl with respect to
various parameters possibly relevant for the observed dispersions:
Width of the central part of the Ti~$3d$ spectral weight dispersion
along $b$ ($w_b^{\prime}$), exchange constant $J$ obtained from the
magnetic susceptibility,\cite{Ruckamp05} hopping probability $t$
derived from $J$, and the hopping integral $t$ of downfolding LDA+U
studies.\cite{Lemmens05}}
\begin{ruledtabular}
\begin{tabular}{lrrr}
                                & TiOBr     & TiOCl    & $\Delta$x/x\\
  \hline
  1D dispersion $w_b^{\prime}$        & 0.13\,eV  & 0.17\,eV & -23\%\\
  J from magnetic susceptibility& 32\,meV   & 58\,meV  & -45\%\\
  $t=\sqrt{J U}/2$              & 0.16\,eV  & 0.22\,eV & -26\%\\
  $t$ from LDA+U                & 0.17\,eV  & 0.21\,eV & -19\%\\
\end{tabular}
\end{ruledtabular}
\end{table}
On the contrary, the experimental width $w_b^{\prime}$ nicely
matches the transfer integral as obtained from $t=\sqrt{J U}/2$
which is smaller in the bromide by $26$\% compared to the chloride.
A similarly fair agreement is achieved with $t$ from the LDA+U
downfolding studies\cite{Lemmens05} where the effective
$d_{xy}$-$d_{xy}$ hopping parameter along the $b$~axis is by $19$\%
smaller in TiOBr with respect to TiOCl. From this analysis it
follows that the experimentally observed dispersions scale with $t$
--- not with $J$ --- and thus can clearly be identified as charge
excitations. Moreover, there is no indication in the spectra of
Fig.~\ref{Figure3} for any asymmetry towards lower binding energies
which could be interpreted as a remnant of the spinon branch. Thus,
we are led to rule out that the experimentally observed spectral
dispersion is a superposition of both holon and spinon excitations.

Turning back to our starting point, {\it i.e.}, the single-band 1D
Hubbard model, a simple explanation for the lack of a spinon
branch and shadow band would be that the two-dimensional coupling
in these compounds under the particularities of a triangular
lattice is already large enough so that generic 1D features of the
spectral weight distribution can not persist. One would then be
left with a lower Hubbard band which though incoherent in nature
could still display sizable dispersion alike to what we observe
experimentally. Clearly, this issue demands for more detailed
theoretical investigation. Alternatively, one could stick to the
1D Hubbard model and cite the coupling to phonons, multiorbital or
spin-Peierls fluctuation effects as possible causes for the
complete suppression of the generic 1D phenomenology. However,
from the organic quasi-one-dimensional conductor TTF-TCNQ we know
that the phonons,\cite{Claessen02,Sing03} which should couple
equally strong to the electrons in this charge density wave system
and are of even higher energy, do not suppress or smear out the
spectral features of spin-charge separation and shadow band.
Moreover, since as already stated the orbital degrees of freedom
are quenched in the oxyhalides and spin-Peierls fluctuations
should not be too important so far above the transition
temperatures, these two effects are probably not very effective.

In view of the above results we are thus left in a situation where
on the one hand the spin-Peierls ground state of both oxyhalides
studied here is dominated by 1D interactions while the analysis and
discussion of the electronic dispersions at room temperature as well
as the incommensurate order in the intermediate phase point to the
importance of 2D (frustrated) interchain interactions. It remains
interesting to see whether or not an anisotropic Hubbard model type
description taking account of the magnetic interchain frustrations
on the underlying triangular lattice is capable of better describing
the electronic properties of this class of materials positioned in
the regime between 1D and 2D correlations.

\section{Conclusions}

In summary, we have investigated the electronic structure of TiOBr
and TiOCl by means of photoelectron spectroscopy and GGA+U
calculations. In these compounds we find a very similar density of
states, with the exception of an enlarged separation between the
manifold of O and Cl(Br) levels and the Ti $3d$ states. This
difference is due to the enlarged unit cell in TiOBr. ARPES
measurements reveal that the Ti~$3d$ spectral weight dispersions of
the two compounds show a great qualitative resemblance, with the
main difference that in TiOBr the overall dispersion width is larger
along the $a$~axis and smaller along the $b$~axis, compared to the
chloride. Thus the quasi-one-dimensional character of TiOBr is
clearly less pronounced. From the comparison of TiOBr and TiOCl, we
are able to show that the dispersion of the $d$ band along the
$b$~axis scales with the square root of the exchange constant $J$,
{\it i.e.} with the hopping integral $t$. Based on this analysis and
the lack of any sign for spin-charge separation or shadow band and
against the background of other experimental findings we argue that
the physics of both systems is governed by the interplay of both
one-dimensional and two-dimensional correlations.
\begin{acknowledgments}
We are grateful to H. Benthien and M. Potthoff for fruitful
discussions. This work was supported by the Deutsche
Forschungsgemeinschaft through SFB 484 and grant CL 124/3-3.
\end{acknowledgments}

\bibliography{TiOBr-references}

\begin{thebibliography}{24}
\bibitem[{foo()}]{footnote-ideal2D}
\bibinfo{note}{We remark that despite $w_b/w_a \approx 1$ in TiOBr it should
  not be considered as almost ideally two-dimensional. Since the physical
  origin of the observed dispersions along $a$ and $b$ and hence their scaling
  properties are not known, their widths cannot be compared offhand.}
\end{thebibliography}

\end{document}